\documentclass[amsmath,superscriptaddress,amssymb,aps,twocolumn]{revtex4}
\usepackage{graphicx}
\usepackage{soul}
\usepackage[colorlinks=true,citecolor=blue,linkcolor=magenta]{hyperref}
\usepackage[usenames]{color}

\def \av#1{{\langle#1\rangle}}

\def \ket#1{|#1\rangle}

\def \be{\begin{equation}}
\def \ee{\end{equation}}
\def \ba{\begin{array}}
\def \ea{\end{array}}
\def \bea{\begin{eqnarray}}
\def \eea{\end{eqnarray}}
\def \nn{\nonumber}

\def \half{{1\over 2}}

\def \a{{\alpha}}

\def \D{{\Delta}}

\def \w{{\omega}}

\def \f{{\varphi}}

\def \yd{^\dagger}
\def \av#1{{\langle#1\rangle}}

\newcommand{\ssm}{\scriptscriptstyle\rm}
\renewcommand{\phi}{\varphi}
\newcommand{\pdag}{\phantom{\dag}}

\begin{document}

\title{Many-body Landau-Zener dynamics in coupled 1D Bose liquids}
\author{Yu-Ao Chen}
\thanks{These authors contribute equally to this work}
\affiliation{Fakult\"at f\"ur Physik, Ludwig-Maximilian-Universit\"at, Schellingstrasse 4, 80798 M\"unchen, Germany}
\affiliation{Max-Planck-Institut f\"ur Quantenoptik, Hans-Kopfermann-Strasse 1, 85748 Garching, Germany }
\affiliation{Institut f\"ur Physik, Johannes Gutenberg-Universit\"at, Staudingerweg 7, 54099 Mainz, Germany}
\author{Sebastian D. Huber}
\thanks{These authors contribute equally to this work}
\affiliation{Department of Condensed Matter Physics, The Weizmann Institute of Science, Rehovot, 76100, Israel}
\author{Stefan Trotzky} 
\affiliation{Fakult\"at f\"ur Physik, Ludwig-Maximilian-Universit\"at, Schellingstrasse 4, 80798 M\"unchen, Germany}
\affiliation{Max-Planck-Institut f\"ur Quantenoptik, Hans-Kopfermann-Strasse 1, 85748 Garching, Germany }
\affiliation{Institut f\"ur Physik, Johannes Gutenberg-Universit\"at, Staudingerweg 7, 54099 Mainz, Germany}
\author{Immanuel Bloch}
\affiliation{Fakult\"at f\"ur Physik, Ludwig-Maximilian-Universit\"at, Schellingstrasse 4, 80798 M\"unchen, Germany}
\affiliation{Max-Planck-Institut f\"ur Quantenoptik, Hans-Kopfermann-Strasse 1, 85748 Garching, Germany }
\affiliation{Institut f\"ur Physik, Johannes Gutenberg-Universit\"at, Staudingerweg 7, 54099 Mainz, Germany}
\author{Ehud Altman}
\affiliation{Department of Condensed Matter Physics, The Weizmann Institute of Science, Rehovot, 76100, Israel}

\date{\today}

\begin{abstract}
The Landau-Zener model of a quantum mechanical two-level system driven with a linearly time dependent detuning has served over decades as a textbook paradigm of quantum dynamics. In their seminal work [L. D. Landau, Physik. Z. Sowjet. \textbf{2}, 46 (1932); C. Zener, Proc. Royal Soc. London \textbf{137}, 696 (1932)], Landau and Zener derived a non-perturbative prediction for the transition probability between two states, which often serves as a reference point for the analysis of more complex systems. A particularly intriguing question is whether that framework can be extended to describe many-body quantum dynamics. Here we report an experimental and theoretical study of a system of ultracold atoms, offering a direct many-body generalization of the Landau-Zener problem.  In a system of pairwise tunnel-coupled 1D Bose liquids we show how tuning the correlations of the 1D gases, the tunnel coupling between the tubes and the inter-tube interactions strongly modify the original Landau-Zener picture. The results are explained using a mean-field description of the inter-tube condensate wave-function, coupled to the low-energy phonons of the 1D Bose liquid.
\end{abstract}

\maketitle

Systems of ultracold bosonic atoms in optical lattices provide clean and highly tunable laboratories for investigating strongly correlated quantum states in low dimensions. Prominent examples in one dimension are the conversion of weakly interacting superfluids into strongly correlated Tonks-Girardeau gases \cite{Paredes04,Kinoshita04,Syassen08,Haller09} or Mott insulating states \cite{Stoferle04} including the measurement of hallmark correlations \cite{Laburthe-Tolra04,Kinoshita:2005} and transport properties \cite{Fertig04}.

Next to the investigation of equilibrium phases, current ultracold atomic systems offer an entirely new perspective on low-dimensional quantum liquids, allowing one to focus on their non-equilibrium dynamical behaviour. Compared to the equilibrium properties, the theoretical understanding of dynamics in such systems is far more rudimentary and is currently a topic of intense study \cite{Cardy06,Kollath07,Manmana07,Rigol08,Gritsev09,Cramer08,Cramer_Flesch08}. A few experiments utilized sudden changes in system parameters in order to induce a quantum quench. This allowed to study fundamental questions including thermalization in low-dimensional systems \cite{Kinoshita06} and interferometry using many-body dynamics \cite{Greiner02a,Hofferberth07,Widera08}.

The limit of adiabatic dynamics, however, is equally interesting and offers a more likely opportunity to observe universal physics \cite{Polkovnikov05a,Zurek05,Polkovnikov08}. The basic prototype of such phenomena is the Landau-Zener (LZ) problem, which describes a sweep through an anti-crossing of two levels \cite{Landau32,zener32}. The two states can be thought of as those of a single particle in two potential wells, which differ in energy by a time-dependent detuning  $\Delta=\alpha \cdot t$ and are tunnel-coupled with strength $J$. The celebrated LZ formula
\begin{equation}
\label{eqn:lz}
P_{\ssm LZ}(J,\alpha) = 1-e^{-2\pi \frac{J^{2}}{\alpha}}
\end{equation}
pertains to the probability for a particle initiated in the bottom well at $t=-\infty$ to end up in the  opposite well at $t=+\infty$, following a linear sweep of the detuning at the rate $\alpha$. The solution of the LZ problem often serves as a starting point for the analysis of more complex problems ranging from sweeps across a Feshbach resonance along the BEC-BCS crossover \cite{Pokrovsky05,Vardi06,Gurarie09} to transport and dissipation in mesoscopic systems \cite{GefenThouless,AoRammer}.

Here we report on an experimental and theoretical investigation of the controlled LZ dynamics in a strongly interacting many-body setting of two tunnel-coupled 1D quantum liquids. Next to a strong dependence of the LZ transition probability on the inter-particle interactions and the condensate fraction, we find a complete breakdown of adiabaticity in the system for the case of a LZ sweep in the excited state, where the transfer fidelity is decreased rather than increased at low sweep rates $\alpha$. We attribute this breakdown to self-trapping of the meta-stable condensate near zero detuning and subsequent decay by phonon emission into the Bose liquid. Strong correlations, which develop in the quantum liquid upon approaching the Mott-insulating phase are found to suppress the self-trapping and decay mechanism, thus restoring the adiabatic transfer.

Our system realizes the controlled LZ dynamics in an array of pairwise coupled potential tubes oriented along the $z$-axis, which are created using a two-dimensional optical lattice potential with a superlattice potential along the orthogonal $x$-direction (see Fig. 1a). The latter is formed by two collinear retro-reflected lattice standing waves with a periodicity ratio of 2 and tunable relative phase \cite{Folling07,Trotzky08}. All lattice depths are given in units of the respective recoil energies $E_r^i=h^2/(2m_{Rb}\lambda_i^2)$  with $m_{Rb}$ being the atomic mass (see Methods). Our experiments begin by loading a BEC of about $9\times10^4$ $^{87}$Rb atoms in the $\vert F=1, m_F=-1\rangle$ state from a magnetic trap into the 2D optical lattice, translating to about $100$ particles in the central tubes (see Methods). At the outset only one tube of each coupled pair, the left tube, is filled with atoms. The remaining empty right tube is far detuned with initial detuning $\Delta_i$, either below ($\Delta_i>0$) or above ($\Delta_i<0$) the left tube. Then the detuning $\Delta$ is changed linearly in time by changing the relative phase of the superlattice, causing particle transfer via tunneling between the tubes. After the sweep, we measure the relative population $n_L$ ($n_R$) in the left (right) tubes respectively, by applying a site-resolving mapping method \cite{sebby07,Trotzky08} (see insets Fig. 1b).

The experimental setup allows us to study how the transfer rate between the tubes depends on the many-body correlations of the 1D Bose gases, and on the ratio between particle interactions and the inter-tube tunnel coupling. The former can be controlled via an additional lattice along the tube axis (the axial $z$-lattice), while the latter can be adjusted via the potential-barrier height set by the depth of short-wavelength lattice along the transverse $x$-direction.

\begin{figure}[t]
\begin{center}
\includegraphics{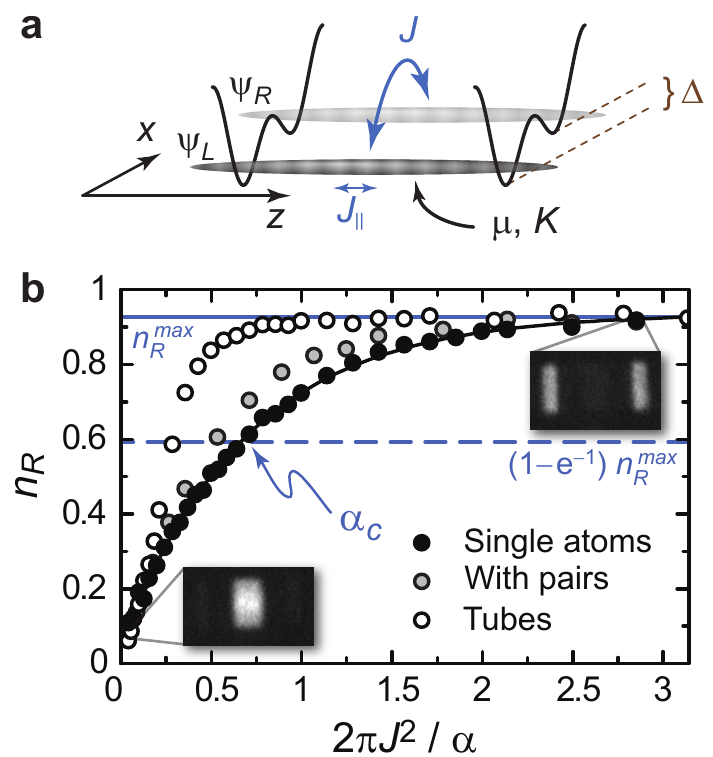}
\end{center}
\caption{Landau-Zener sweeps in a quantum ladder. \textbf{a} Schematic sketch of the experimental setup. Pairs of 1D Bose liquids are tunnel coupled through a potential barrier along the $x$-direction. In this system, the potential offset (bias) $\Delta$ between the tubes, the inter- and intra-tube kinetic energies $J$ and $J_\parallel$, the chemical potential $\mu$ and the correlations properties of the 1D gases characterized by the Luttinger parameter $K$ can be controlled. \textbf{b} Measured relative populations in right tube $n_R$ after ground state sweeps as a function of the inverse sweep rate $2\pi J^2/\alpha$ with single atoms (black circles), single atoms and pairs (grey circles) and with 1D Bose liquids (open circles). The insets show typical band mapping images from which the populations in the left and right well are deduced.}\label{fig:dot}
\end{figure}

\section{Ground-state sweep}

We begin our discussion with the case of ground-state sweeps, where $\Delta_i<0$ such that the filled left tube is the one with lower energy. The comparison of transfer efficiencies ($n_R$) for pairwise coupled dot-like lattice sites and one-dimensional tubes as a function of sweep rate is shown in Fig. 1b. The former scenario is realized by adding a deep lattice along the axial $z$-direction. To quantify the sweep fidelity with a single number, we define the characteristic  rate $\alpha_c$ as the sweep rate to achieve a transfer efficiency of $1-\mbox{e}^{-1}\approx 63\%$ with respect to the maximal measured value. Therefore, larger $\alpha_c$ correspond to higher transfer rates. In the case of single atoms in the dot-like sites, the result almost matches with the LZ formula Eq.\,(\ref{eqn:lz}). Here, $\alpha_c = 2\pi J^2$ sets the natural scale for the sweep rate. If the filling is increased, we find an enhanced value of $\alpha_c$. This enhancement is even more pronounced in the case of the pairwise coupled tubes filled with up to 100 atoms (see Fig. 1b).

The increase of transfer rate with atom number can already be understood within the simple two-mode model of a ``double dot" occupied by strongly interacting particles \cite{Cheinet08,Venumadhav10}.  For the case of strong interactions within a mode compared to the tunnel coupling, the LZ transition is split into $N$ avoided level crossings, each corresponding to the transfer of one particle with a coupling enhanced by Bose statistics (see Supplementary Information).
Note that interactions are crucial to observe this enhancement: a non-interacting BEC would result in the same transfer rate as is the case for a single particle.

From now on we focus on the the case of large particle numbers. Then an effective 1D Bose liquid forms in the filled tubes, and we study how the transfer efficiency to the empty tubes depends on the properties of that liquid. We control the nature of the many-body system by tuning the strength of both the transverse $x$-lattice and axial $z$-lattice potentials. Both parameters strongly influence the characteristic sweep rate $\alpha_c$ in the LZ problem (see Fig. 2): $\alpha_c$ is found to increase with increasing strength of the transverse $x$-lattice potential while it decreases when increasing the axial $z$-lattice depth. For very large $z$-lattice depth, the system is eventually driven across the Mott transition in the ladder system \cite{danshita07} and deep in the Mott-insulating regime for unity filling, the single-particle result is restored.

\begin{figure}[t]
\begin{center}
\includegraphics{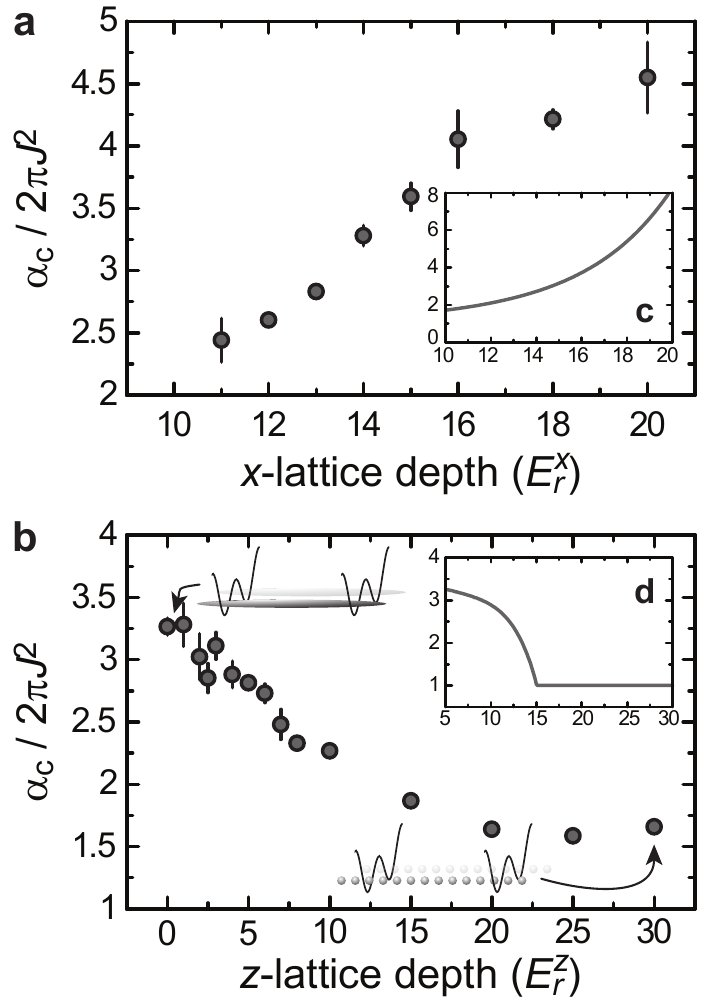}
\end{center}
\caption{Influence of interactions and correlations on the characteristic sweep rate. Experimentally measured scaled characteristic sweep rate $\alpha_c/(2\pi J^2)$ as a function of the transverse (\textbf{a}) and axial lattice depth (\textbf{b}). The insets (\textbf{c,d}) show the theoretical predictions using mean-field theory (Eq.\,(\ref{Hmf})) as described in the text.}
\label{fig:forward}
\end{figure}

In order to explain this behaviour, we consider an infinite homogeneous Bose-Hubbard ladder with an average of one particle per rung (see Methods). We treat the inter-tube coherence by a mean-field approximation. That is, we take the quasi-condensate in the ladder to be in a fixed linear combination of left and right tube $\hat a^\dagger_i=\psi_L \hat b^\dagger_{Li}+\psi_R \hat b^\dagger_{Ri}$. This leads to a reduced single-chain Hamiltonian $H_{\ssm eff}[\psi_L,\psi_R]$, which depends on the variational parameters $\psi_L$ and $\psi_R$ (Eq.\,(\ref{Heff}) in Methods).  Minimization of the ground-state energy $E(\psi_L,\psi_R)$, subject to the constraint $|\psi_L|^2+|\psi_R|^2=1$ defines a non-linear LZ problem, that is analogous to the two-mode non-linear Sch\"odinger equation in ref. \cite{Wu00,Liu02}.

To solve for $E(\psi_L,\psi_R)$ we employ a mean-field theory of the single-chain Hamiltonian (see Eq. (\ref{Heff}) in Methods), which leads to decoupled local Hamiltonians on rungs of the ladder:
\begin{eqnarray}
H_{\ssm MF}&&=- J_{\parallel}(\Phi \hat a^\dagger_i+\Phi^* \hat a^{\vphantom{\dagger}}_i)+U(|\psi_L|^4+|\psi_R|^4)\hat{a}^\dagger_i \hat{a}^\dagger_i \hat{a}^{\vphantom{\dagger}}_i \hat{a}^{\vphantom{\dagger}}_i \nonumber\\
&&+\left[\Delta(|\psi_L|^2-|\psi_R|^2)-J(\psi^*_L\psi_R+\text{c.c.})\right]\hat{a}^\dagger_i \hat{a}^{\vphantom{\dagger}}_i.
\label{Hmf}
\end{eqnarray}
Here $\Phi=\av{\hat a_i}$ is the mean-field condensate. While this approximation does not capture the subtle long-range correlations of the one-dimensional system, it is rather good for the energetics. In the weak coupling limit $\av{\hat a_{i}}\approx\sqrt{n}$, and Eq.\,(\ref{Hmf}) reduces to the usual two-site Gross-Pitaevskii functional \cite{Wu00}. At stronger interactions, the condensate $\Phi$ is depleted until it vanishes at the transition to the Mott-insulating state in the ladder.

Within the mean-field  Eq.~(\ref{Hmf}), the time dependence  of the two modes $\psi_L$ and $\psi_R$ is governed by a non-linear Schr\"odinger equation similar to ref.\, \cite{Wu00,Liu02}, with the effective non-linearity $\eta\approx U\Phi^2/J$ (see Supplementary Information). We can therefore apply an adiabatic perturbation theory \cite{Liu02}, in order to obtain a modified LZ formula $P_R=1-\exp(-2\pi q J^2/\alpha)$ with an effective coupling strength $\sqrt{q}J$. Here $\alpha_c=2\pi q J^2$, with $q$ given in ref. \cite{Liu02} and in the Supplementary Information as a function of the effective non-linearity $\eta$.

The characteristic rates $\alpha_c$ calculated in this way are shown as insets in Fig. 2. Their dependence on the transverse $x$-lattice and the axial $z$-lattice depth accounts well for the trends observed in the experiments. Both of these trends stem from the change of the effective non-linearity $\eta$. An increase in the potential barrier between the tubes ($x$-lattice depth) increases $\eta$ and reduces the curvature of the ground-state energy in its dependence on detuning, thus allowing for faster sweeps \cite{Arimondo} (corresponding ultimately to the split resonance case discussed above for $\eta \gg 1$).  Increasing the $z$-lattice depth along the tubes first increases $\eta$ via an increase in the effective $U$. For even larger lattice depths, however, the condensate is significantly depleted when driven towards the Mott transition of the quantum ladder. Deep in the Mott regime all Bose enhancement in the tunneling process is lost and one essentially recovers the single-particle LZ results. The fact that the calculated $\alpha_c$ goes to 1 immediately at the transition is an artifact of the mean-field theory (Eq. (\ref{Hmf})). Quantum fluctuations beyond mean-field theory give rise to on-site particle number fluctuations and coupling between the sites even in the Mott phase \cite{Gerbier07}. This will smoothen out the change of $\alpha_c$ across the transition, saturating to the asymptotic value $\alpha_c/2\pi J^2=1$ deep in the Mott-insulating regime. We attribute the slightly higher saturation value seen in the experiment ($\alpha_c \approx 1.5$) to the fraction of doubly occupied sites that are present in the system.

\section{Inverse sweep}
\begin{figure}[ht]
\begin{center}
\includegraphics{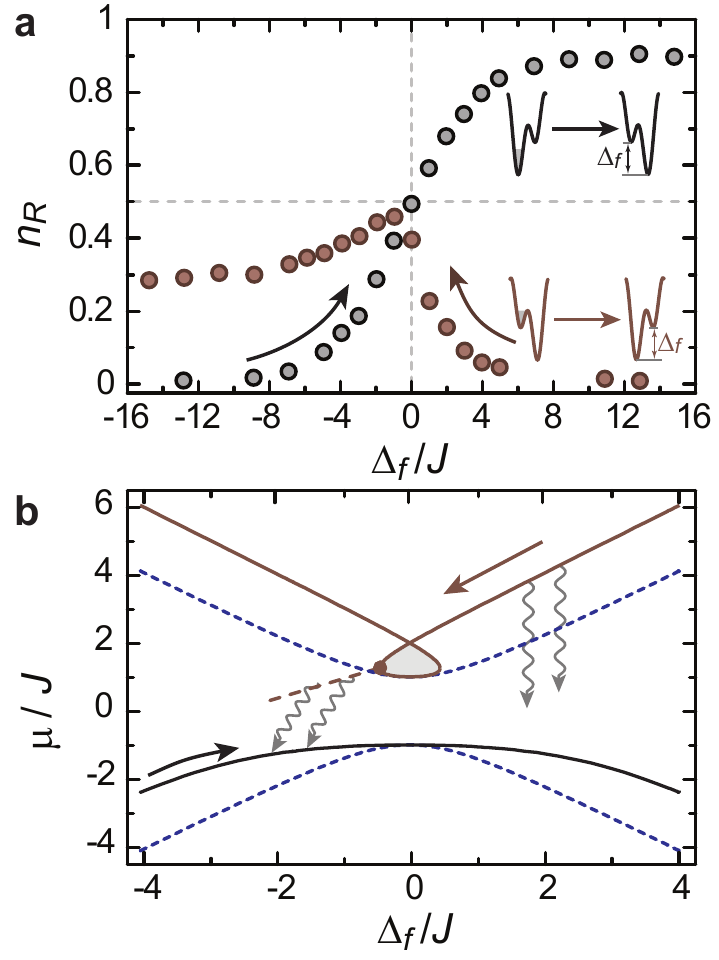}
\end{center}
\caption{Adiabaticity breakdown in the inverse LZ sweep. \textbf{a} Comparison of the transfer efficiency $n_R$ from the filled to the initially empty tube as a function of the final detuning for a ground state sweep (black dots) and an inverse sweep (red dots). The sweeps were carried out at a constant rate $2\pi J^2/|\alpha|=2.1(1)$ and $x$-lattice depth of  15 $E_r^x$. For the inverse sweep a sudden breakdown in transfer fidelity is observed beyond resonance. \textbf{b} Adiabatic energy levels for ground state sweep (black curve) and inverse sweep (red curve). For strong repulsive interactions a loop develops in the condensate energy level structure \cite{Wu00}. In an inverse sweep the condensate becomes self-trapped as it follows the path along the upper branch until reaching the spinodal point in the outer edge of the loop (red dot). Beyond this point it becomes unstable and decays to the lower branch, while emitting phonon excitations along its axial direction. The dashed blue curve denotes the energy levels of the non-interacting problem.}
\label{fig:loop}
\end{figure}

We now turn to the investigation of the inverse LZ sweep. Namely, we initialize the system with $\Delta_i>0$ such that the filled left tube is the one with higher energy. We follow the population of the tubes as the detuning is gradually changed until the relative potential between the tubes is reversed.

In Fig. 3a we plot $n_R$ as a function of the final detuning $\Delta_f$ for a relatively slow sweep with $2\pi J^2/|\alpha|=2.1(1)$. While the ground-state sweep shows a continuous transfer of population, the inverse sweep exhibits a slightly reduced transfer for $\Delta_f>0$ and an abrupt breakdown for $\Delta_f<0$ once the resonance is crossed. From then on, particles are transferred back to the left tube which is now lower in energy. We will argue below that the dramatic difference between the two cases stems from the highly non-equilibrium nature of the inverse sweep compared to the near-equilibrium conditions of the ground-state sweep. In Fig. 3b we illustrate the two mechanisms responsible for the breakdown of adiabaticity: 1) axial phonons which provide a relaxation mechanism and 2) a loop structure of the mean-field condensate energy, which leads to self-trapping.

\begin{figure}[t]
\begin{center}
\includegraphics{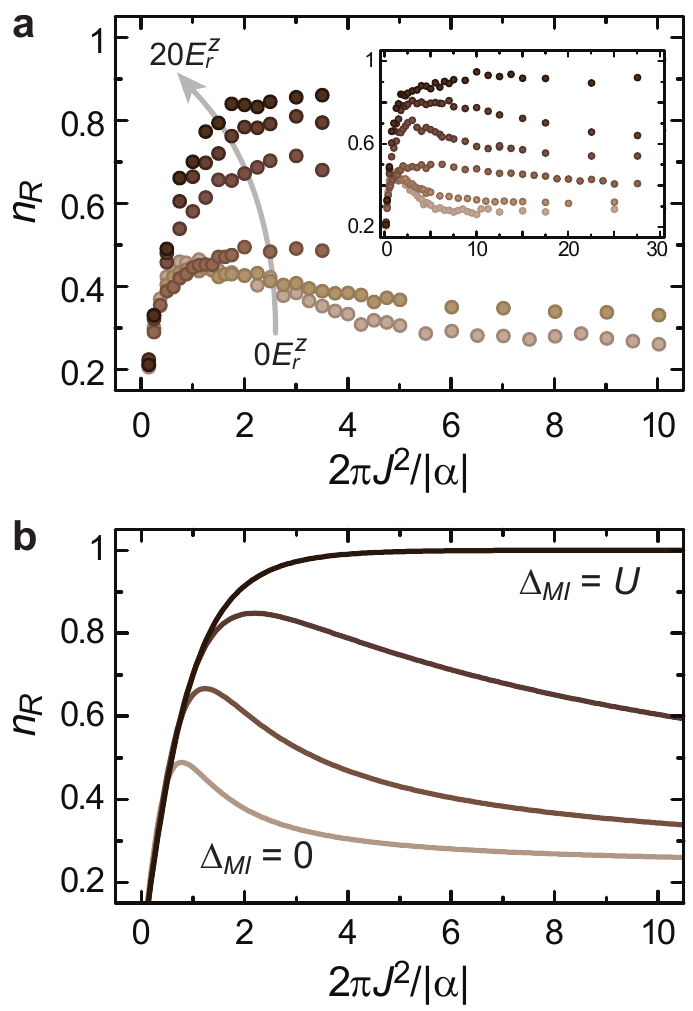}
\end{center}
\caption{Influence of intra-tube correlations on the adiabaticity breakdown. \textbf{a} Measured transfer efficiency $n_R$ as a function of the scaled inverse sweep rate for different axial $z$-lattice depths. The $x$-lattice depth was 12 $E_r^x$, and the $z$-lattice depths were set to 0 $E_r^z$, 2 $E_r^z$, 4 $E_r^z$, 7 $E_r^z$, 10 $E_r^z$ and 20 $E_r^z$, respectively. The inset shows the transfer efficiency extended to slower sweeps. \textbf{b} Transfer probability as a function of scaled inverse sweep rate computed using an effective Luttinger liquid theory (see Eqs.\,(\ref{eqn:powerlaw},\ref{eqn:decayprob})) valid at weak non-linearity (see text).}
\label{fig:panels}
\end{figure}

Fig. 4a shows a more detailed study of the adiabaticity breakdown for different axial $z$-lattice depths. We see that at fast sweeps ($2\pi J^2/|\alpha| < 1$), $n_R$ increases with decreasing sweep rate as expected from the standard LZ problem. In this regime, the curves corresponding to different $z$-lattice depths follow the typical universal scaling with $2\pi J^2/|\alpha|$. By contrast, in the slow sweep regime $2\pi J^2/|\alpha|> 1$, this scaling is no longer obeyed. Moreover, we find the transfer efficiency $n_R$ to actually decrease with decreasing sweep rate for sufficiently low $z$-lattice depths, indicating a complete breakdown of adiabaticity.

Such a decay can be understood as follows. Particles occupying the upper state close to $\Delta=0$, i.e. the high-energy superposition of left and right tubes, form a filled 1D Bose liquid. The orthogonal low-energy superposition state forms an effective empty mode into which particles or pairs can decay spontaneously. The respective decay rates $\Gamma_1$ and $\Gamma_2$ depend on the couplings set by the interaction constant, and on the correlations in the Bose liquids.  To estimate $\Gamma_1$ and $\Gamma_2$ we resort to the universal low-energy description of the Bose liquid in the filled mode in terms of a Luttinger liquid \cite{Haldane81}. Within Fermi's golden rule, we find the decay rates to be \cite{Huber09}:
\begin{eqnarray}
\label{eqn:powerlaw}
\Gamma_1(\Delta,K)&=&U\lambda_1^2 F_{1}(K)\left(\frac{\mu}{\Delta_{\ssm eff}}\right)^{1-1/2K^*},\nonumber\\
\Gamma_2(\Delta,K)&=& U\lambda_2^2 F_{2}(K)\left(\frac{\mu}{\Delta_{\ssm eff}}\right)^{1/2-2/K^*}.
\label{rates}
\end{eqnarray}
Here $\lambda_1=2\Delta J/\Delta_{\ssm eff}^2$, $\lambda_2=4J^2/\Delta_{\ssm eff}^2$, relate to single-particle and pair tunneling between the upper and lower states, respectively, while $\Delta_{\ssm eff}=\sqrt{4J^2+\Delta^2}$ denotes the actual energy difference between these states. The functions $F_{1}(K)\approx 4\pi$ and $F_2(K)\approx \pi/\sqrt{8}$ are very weakly dependent on the Luttinger parameter $K$ and $K^*=K/(1+\delta_s)^{2}$ is an effective Luttinger parameter, renormalized by a non-universal phase shift due to the interaction between the quantum liquid and the particle or pair transferred to the empty lower state (See ref. \cite{Castro-Neto96} and Supplementary Information).

If we assume random independent decay events, then the density in the upper state obeys the equation ${\dot\rho}=-(\Gamma_1+2\Gamma_2)\rho^2\equiv -\Gamma\rho^2$. The decay is significant as long as the detuning between the tubes is less than the chemical potential set by the interactions in the filled tube. At larger detuning, which exceeds the linear phonon spectrum in that tube, the decay process is greatly suppressed due to phase space limitations.

The density remaining in the upper state can now be found from the solution of the decay equation with the rates in Eq.\,(\ref{rates})
\begin{eqnarray}
\rho=\rho_0\left[1+\frac{2\rho_0}{|\alpha|}\,\int_0^{\mu} d\Delta\,\Gamma(\Delta)\right]^{-1}.
\label{eqn:decayprob}
\end{eqnarray}
This result is valid in the limit of slow sweeps, when the time spent in the decay zone $t_d=2\mu/\alpha$ is longer than the shortest response time of the Luttinger liquid $\simeq h/\mu$. For shorter sweep times, the response is essentially that of free particles. Then the single-particle LZ formula Eq.\,(\ref{eqn:lz}) is expected to apply.

The above arguments can be extended to the Mott insulator, where the opening of the gap in the excitation spectrum gradually eliminates the decay channels. This is taken into account by introducing the Mott gap $\Delta_{\ssm MI}$ as a lower cutoff to the integral over $\Gamma$ in Eq.~(\ref{eqn:decayprob}). Deep in the Mott phase the decay vanishes completely as the system becomes a collection of effectively decoupled double wells.

Taking all these points into account, we plot the expected transfer efficiency $n_R$ as a function of the inverse sweep rate $2\pi J^2/|\alpha|$ in Fig.~4b. The results have been calculated for different values of the axial $z$-lattice depth, taking into account the inhomogeneous filling of different tube pairs in the experiment (see Methods). Both experiment and theory show the same qualitative behavior.

\begin{figure}[b]
\begin{center}
\includegraphics{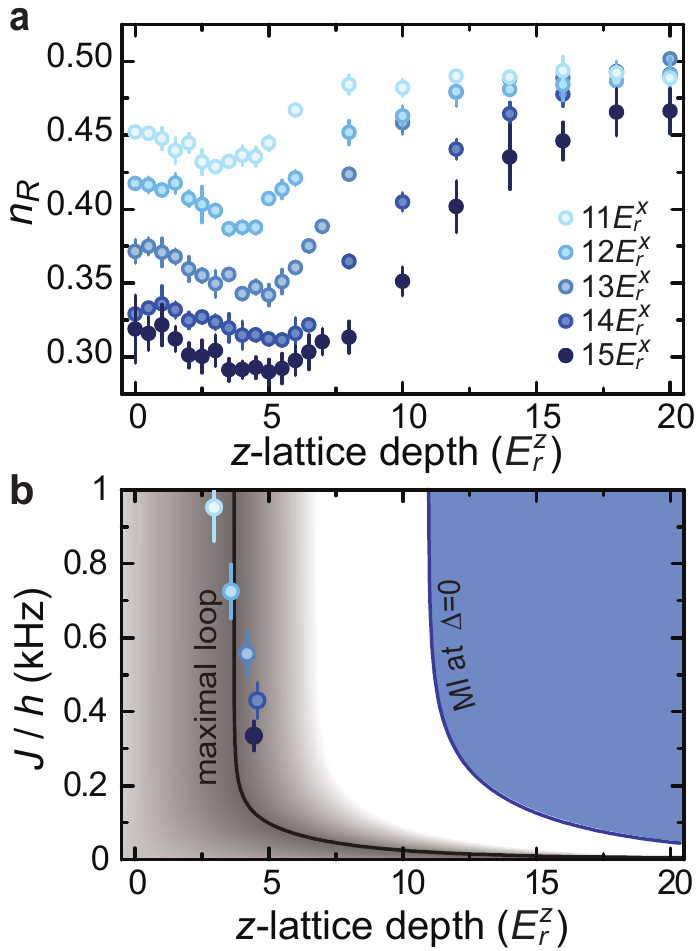}
\end{center}
\caption{Changing the loop size: influence of intra-tube correlations on transfer efficiency. \textbf{a} Measured transfer efficiency at a constant normalized sweep time $2 \pi J^2/\alpha=0.53(3)$ as a function of the axial $z$-lattice depth and different inter-tube tunnel couplings. A pronounced minimum of the transfer efficiency is observed prior to reaching the Mott transition in the quantum ladder. This minimum shifts towards higher $z$-lattice depths for decreasing inter-tube couplings $J$. \textbf{b} Phase diagram of the meta-stable upper condensate branch computed using mean-field theory (Eq.\,(\ref{Hmf})). In the gray shaded area, the upper condensate branch exhibits a loop (Fig.~3b). The datapoints represent the minima in the measured transfer efficiency as obtained from a fourth-order polynomial fit to the data in \textbf{a}. They approximately coincide with the theoretical prediction of the maximal loop size (maximal non-linearity in the mean-field picture).}
\label{fig:5}
\end{figure}

In addition to the breakdown of adiabaticity, we observe an intriguing non-monotonous change in the transfer efficiency versus the axial $z$-lattice depth in the crossover between the slow and fast sweep regimes ($0.5<2\pi J^2/\vert\alpha\vert<1$) (see Fig. 4a). The transfer efficiency $n_R$ first decreases and then increases with increasing $z$-lattice depth. A detailed investigation of this effect is plotted in Fig.~5a, where we show how $n_R$ changes at a fixed value of $2\pi J^2/|\alpha|$ with increasing $z$-lattice depth as the system evolves from a superfluid to a Mott insulator. The experimental data shows a pronounced minimum of $n_R$ prior to reaching the Mott transition.

In order to explain this phenomenon, the loop structure, caused by the non-linearity and plotted in Fig.~3b, needs to be taken into account. Let us consider the mean-field Hamiltonian (Eq.\,(\ref{Hmf})) in the weak coupling regime, far from the Mott phase. Extremizing the mean-field energy with respect to the inter-tube coherences leads to a non-linear version of the two state Schr\"odinger equation (see Supplementary Information). For a non-linearity $\eta >1$, the ``energy levels'' versus detuning develop a loop structure \cite{Wu00}, as seen in Fig.~3b, which grows in width with increasing non-linearity.

Such a loop structure leads to a dramatic dynamical evolution of the condensate once the resonance is crossed in the inverse sweep. First the condensate starts out far detuned on the right upper branch, which corresponds to a local extremum of the mean-field energy. In the course of a slow sweep the condensate moves along this branch as the detuning is decreased, and gets caught in the loop structure in a self trapped state. In a one-dimensional system such a state survives as a slowly decaying meta-stable state \cite{Hipolito}. The edge of the loop is a spinodal point of the mean-field energy where the local extremal solution disappears. When the detuning is decreased further, the condensate finds itself in a catastrophic scenario on a downward slope of the energy landscape, leading to a decay to the low-energy mode of the system via the discussed emission of phonons into the 1D quantum liquid (see Supplementary Information).

In our case, the non-linearity $\eta$ and therefore the size of the loop structure depends non-monotonously on the axial $z$-lattice depth. Far from the Mott transition the lattice acts to increase the effective on-site interaction and with it the effective non-linearity $\eta$, which leads to growing loop size  and enhanced probability of the condensate to be self-trapped. As a consequence, $n_R$ first decreases with increasing $z$-lattice depth (see Fig.~5a). However, beyond a certain lattice depth the condensate $\Phi$ begins to deplete significantly, causing a decrease of $\eta$ and hence an increase of $n_R$ until $\eta$ vanishes close to the Mott transition where $\Phi=0$. Fig.~5b charts the line of maximal non-linearity and the Mott phase boundary as a function of $z$-lattice depth and the inter-tube tunneling $J$. The former agrees well with the experimentally determined position of the minima in transfer efficiency seen in Fig.~5a. Note that the decay mechanism still plays a role until deep in the Mott regime. Then, however, the phonon decay channel is shut off by the Mott gap and thus we recover the standard single-particle LZ behavior.

Finally, let us comment on the probability for the system to become self-trapped in the loop structure for different sweep rates. For very fast sweeps the behavior of the non-linear system is close to the linear one as the system is unlikely to be trapped in the loop structure. However, for slower sweeps and larger non-linearities (leading to larger loop structures) the probability for self-trapping is strongly enhanced. We find that the crossover between the slow and fast sweep regimes, for strong non-linearity, takes place at inverse sweep rate $J^2/\alpha_\star\approx \eta^{-2/3}$  \cite{Liu02}.

\section{Discussion and conclusions}

We have measured and analyzed novel dynamical phenomena in pairwise coupled 1D Bose liquids, which constitute a direct generalization of the LZ dynamics in a many-body system. A particularly striking result is the sudden breakdown of adiabaticity seen in the inverse sweep, for which slower sweeps lead to lower transfer efficiency. We explained the results theoretically by treating the inter-tube coherence within a mean-field approximation valid all the way to the strongly correlated regime. The analysis lead to an effective non-linear LZ problem similar to the one described in ref. \cite{Wu00,Liu02}. The breakdown of adiabaticity in the inverse LZ sweep can then be attributed to self-trapping of the inter-tube coherence in a meta-stable solution, associated with a loop structure in the energy levels, which emerges beyond a critical value of the non-linearity. We showed that the effective non-linearity decreases and eventually vanishes upon approaching the transition to the Mott-insulating phase. A crucial new ingredient in the present setting is the coupling of the inter-tube coherence to the low energy excitations (phonons) of the 1D Bose liquid. These fluctuations, treated within a quantum Luttinger liquid theory, provide the necessary dissipation mechanism, which allows for decay of the condensate from the meta-stable solution.

Our work provides new insight into the quantum dynamics of many-body systems far from equilibrium, and constitutes a step towards extending the LZ framework for treating such phenomena. An interesting scenario arises when our approach is applied to fermionic systems. The presence of an underlying Fermi surface can, in particular, facilitate the observation of universal phenomena in the non-equilibrium quantum dynamics, such as predicted for the Rabi oscillations in a similar setup consisting of a coupled tubes undergoing a sudden quench \cite{Huber09}. Another natural continuation of our work is to the dynamics of quantum spins in an optical lattice, for which the possibility of universal phenomena was suggested \cite{Barmettler09}.

\section*{Methods}

\textbf{Experimental cycle.} We create the array of pairwise coupled tubes by means of a bichromatic superlattice introduced in ref. \cite{Folling07}. The atoms are first loaded into a two-dimensional optical lattice with periodicity of $\lambda_s/2 = 765\,{\rm nm}$ along the $x$-direction and $\lambda_y/2=420\,{\rm nm}$ along the $y$-direction. This is achieved by ramping up the lattices to $V_s = 40\,E_r^s$ and $V_y = 30\,E_r^y$, respectively, within $300\,{\rm ms}$. In this way, we form a 2D array of tubes with about 100 particles in the central tubes. A short-period lattice along the $x$-direction with periodicity $\lambda_x/2=382.5\,{\rm nm}$ allows us to split the $x$-lattice sites into double-wells. By ramping up this lattice to $50\,E_r^x$ within $10\,{\rm ms}$ with the relative phase set to obtain highly tilted double-wells, we achieve loading of all atoms to left tubes alone. Any coupling in the $y$-direction is frozen out by ramping up the $y$-lattice to $50\,E_r^y$ within $100\,{\rm ms}$. During this ramp time, the $z$-lattice with wavelength $\lambda_z=840$ nm is ramped to the desired depth to tune the intra-tube tunneling $J_\parallel$ and the relative phase between $x$- and $s$-lattice is adjusted to the initial detuning $\Delta_i= E_L - E_R$, with $E_L$ ($E_R$) the energy of the left (right) well ($\Delta_i/h=-5\,{\rm kHz}$ for the ground-state sweeps and $\Delta_i/h=6.6\,{\rm kHz}$ for the inverse sweeps).

Before starting the LZ sweep, we set the inter-tube coupling $J$ by rapidly ramping down the $x$-lattice and thereby the double-well barrier in $200\,{\rm \mu s}$. Then we perform the LZ sweep by linearly changing the detuning between left and right tube to a final value $\Delta_f$ with a sweep time $T = (\Delta_f-\Delta_i)/\alpha$ by changing the relative phase between the $x$- and $s$-lattice ($\Delta_f/h=5\,{\rm kHz}$ for ground state sweep and $\Delta_f/h=-6.6\,{\rm kHz}$ for inverse sweep). After the sweep is finished, we ramp up the barrier in $200\,{\rm \mu s}$ to quench the coupling again. By applying a site resolving mapping method described in Refs. \cite{sebby07,Folling07}, we measure the relative population in left and right tubes, respectively, in order to detect the transfer efficiency of the sweep.

\textbf{Bose-Hubbard ladder model.} The double-tubes, with an optical lattice depth $V_z>5E_R$ along the tube axis, can be described in terms of a Bose-Hubbard ladder model:
\begin{eqnarray}
H&=&-J_\parallel\sum_{\av{ij},\nu=L,R}\left(\hat b^\dagger_{i\nu} \hat b^{\vphantom{\dagger}}_{j\nu}+H.c.\right)\nonumber\\
&&+\sum_{i,\nu=L,R} U \hat b^\dagger_{i\nu} \hat b^\dagger_{i\nu} \hat b^{\vphantom{\dagger}}_{i\nu} \hat b^{\vphantom{\dagger}}_{i\nu}\nonumber\\
&&-\sum_{i} \Delta(\hat n_{i R}-\hat n_{i L})+J(\hat b^\dagger_{i R} \hat b^{\vphantom{\dagger}}_{i L}+\text{H.c.}).
\label{bhm}
\end{eqnarray}
By treating the inter-tube coherence in a mean-field approximation we reduce from a two chain model to an effective one chain model for the quasi condensate mode $\hat a^\dagger_i=\psi_L \hat b^\dagger_{i L}+\psi_R \hat b^\dagger_{i R}$:
\begin{eqnarray}
\label{Heff}
H_{\ssm eff}&=&-J_\parallel\sum_{\av{ij}}\left(\hat a^\dagger_i \hat a^{\vphantom{\dagger}}_j+H.c.\right)\\
&&+\sum_i U(|\psi_L|^4+|\psi_R|^4)\hat a^\dagger_i \hat a^\dagger_i \hat a^{\vphantom{\dagger}}_i \hat a^{\vphantom{\dagger}}_i
\nonumber\\
&&+\sum_i\left[\Delta(|\psi_L|^2-|\psi_R|^2)-J(\psi^*_L\psi_R+c.c.)\right] \hat a^\dagger_i \hat a^{\vphantom{\dagger}}_i.
\nonumber
\end{eqnarray}
This should be viewed as a variational Hamiltonian which defines a non-linear LZ problem for the inter-tube coherences $\psi_L,\psi_R$ as described in the text.

\textbf{Transfer efficiency in inverse sweep.} In order to compare this theory with the measurements we also need to account for the inhomogeneity in the experiment. Due to the transverse parabolic confinement, the experiment actually averages over tubes with different particle numbers. The central tubes have high particle number, while more peripheral tubes have fewer particles. In quite a high fraction of the double-tubes, the chemical potential is lower than the level spacing between phonon states, and these act as zero dimensional double dots where the decay processes are absent. Taking into account those tubes we have the approximate formula for the slow sweep regime $ n_R=(1-\nu_0)P_R+\nu_0 $ where $\nu_0$ is the fraction of the particles in the peripheral tubes. To test this hypothesis we repeated the experiment with varying particle number. The expectation is that $\nu_0$ will decrease with increasing particle number $N$ because more of the tubes will fill up considerably and enter the 1D regime. In the experiment we indeed found that the saturation value of $n_R$, in the limit of slow sweeps, decreased with increasing $N$.

We would like to thank Anatoli Polkovnikov, Giuliano Orso, Christian Kasztelan and Ulrich Schollw\"ock for stimulating discussions. This work was supported by the DFG, the EU (STREP NAMEQUAM), DARPA (OLE program), AFOSR, DIP (EA and IB), and the ISF (EA).


\section*{Supplementary Information}

\renewcommand{\thefigure}{S\arabic{figure}}
 \setcounter{figure}{0}
\renewcommand{\theequation}{S.\arabic{equation}}
 \setcounter{equation}{0}
 \renewcommand{\thesection}{S.\Roman{section}}
\setcounter{section}{0}

\section{Few-particle Landau-Zener sweeps in a double dot}

A simple picture for a multi-particle LZ sweep can be obtained when considering a double dot filled with $N$ interacting particles. For strong repulsive interactions $U\gg J$, the ground state undergoes $N$ well separated (independent) tunneling resonances as the bias $\Delta $ is changed from $-\infty$ to $\infty$ \cite{Cheinet08-s}. At each resonance, two Fock states $|n_L,n_R\rangle$ and $|n_L-1,n_R+1\rangle$ are coupled by single-particle tunneling. The corresponding coupling matrix element is
\begin{eqnarray}
  -J_{n_R}^N &\equiv& -J \langle n_L,n_R| \hat a^\dagger_L \hat a_R |n_L-1, n_R+1\rangle \nonumber\\
             &=& -\sqrt{n_L(n_R+1)} J \,,
\end{eqnarray}
where the emerging prefactor $>1$ stems from the bosonic statistics and $n_L+n_R=N$. We will now consider a linear sweep with $\Delta = \alpha \cdot t$ starting from $\Delta(t=-\infty)=-\infty$ and calculate the overall transfer efficiency from the left to the right well at $t=\infty$. Herein, we will neglect the $k$-th order tunnel couplings ($k\geq 2$) which occur between the excited states and are suppressed as $(J/U)^{k-1}$. The probability to stay adiabatic at a specific first order resonance is given by substitution of Eq.~(S1) into Eq.~(1) as $P_{LZ}(J_{n_R}^N,\alpha)$. The probability to adiabatically follow the ground state throughout the whole sweep thus is
\begin{equation}
  P_{\ssm ad}(\alpha) = \prod_{i=0}^{N-1} P_{LZ}\left(J_{n_R}^N,\alpha\right)\,.
\end{equation}

Taking into account the remaining right-well population at each diabatic crossing of a resonance, we obtain the transfer efficiency $n_R(\alpha)$ of the sweep:
\begin{equation}
  n_R(\alpha) = \prod_{i=0}^{N-1}\left[1- \frac{N-i}{N}\, \left(1-P_{LZ}\left(J_{n_R}^N,\alpha\right)\right)\right]\,.
\end{equation}

\begin{figure}[t]
\begin{center}
\includegraphics{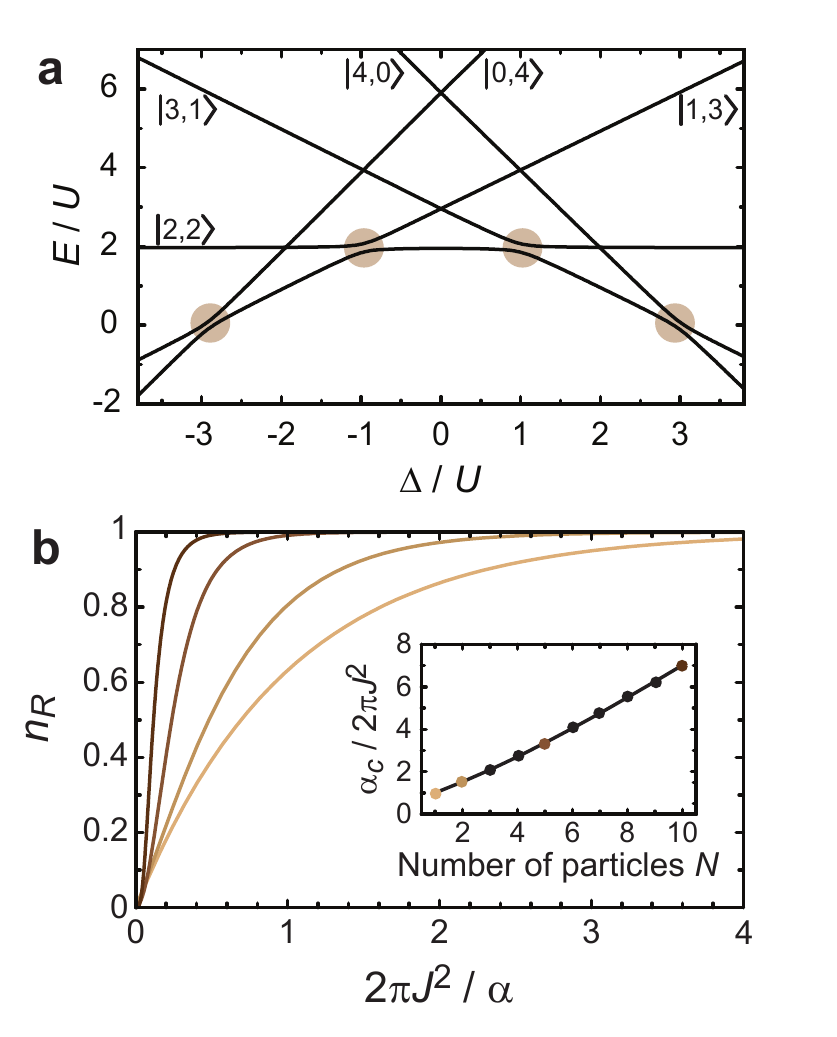}
\end{center}
\caption{
Double-dot picture in the limit of strong interactions. \textbf{a} Plot of the eigenenergies versus $\Delta$ for a double dot filled with four particles and $U/J=30$ The shaded regions mark the individual tunnel resonances crossed in the ground-state LZ sweep. \textbf{b} Plot of the right-well population after a LZ sweep in a double dot with strongly interacting particles ($U\gg J$) and rate $\alpha$ as calculated from Eq. (3). The transfer efficiency increases with increasing number of particles $N$ in the double dot. The inset shows the characteristic sweep rates $\alpha_c$ with $n_R(\alpha_c) = 1-e^{-1}$.
}
\label{fig:largeU}
\end{figure}

In Fig.~\ref{fig:largeU}, we plot $n_R(\alpha)$ for various numbers of particles $N$ in the double dot. The characteristic sweep rate $\alpha_c$ increases gradually with $N$ and we find an enhancement by a factor of 7 for $N=10$.

\begin{figure}[t]
\begin{center}
\includegraphics{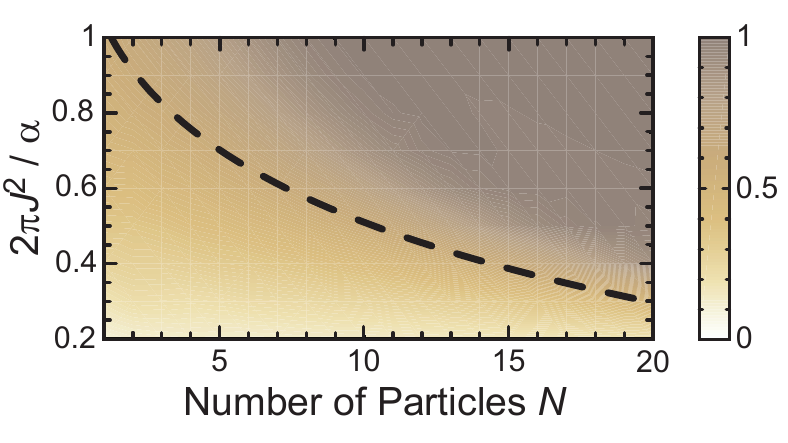}
\end{center}
\caption{
Transfer efficiency for moderate interactions. We plot the simulated transfer efficiency $n_R(\alpha)$ for various numbers particles and an interaction energy of $U = 0.5J$. Here, the tunnel resonances overlap. The enhancement of the characteristic sweep rate $\alpha_c$ (dashed line) with increasing $N$ persists, but becomes less strong compared to the case of independent tunnel resonances.
}
\label{fig:UJ05}
\end{figure}

For smaller interactions $U \sim J$, the tunnel resonances overlap and cannot be treated independently. We have performed numerical simulations of sweeps with $U/J = 0.5$ using exact diagonalization. We plot $n_R(\alpha)$ as obtained from these simulations for $N=1\ldots20$ in Fig.~\ref{fig:UJ05}. The results show the increase of $\alpha_c$ with $N$, where the enhancement for $N=10$ is reduced to a factor of 2 as compared to the case of independent resonances.

We note that the fidelity for an inverse sweep in a double dot is governed by the $N$-th order coupling between the states $|N,0\rangle$ and $|0,N\rangle$ which is suppressed as $(J/U)^{N-1}$, yet finite. Therefore, it is always possible to stay adiabatic for slow sweeps in the double dot system, even though the adiabaticity timescales might become unrealistically long. This is in contrast to the results for pairwise coupled tubes presented in this article, where mean-field and Luttinger-liquid physics lead to a breakdown of adiabaticity for slow sweeps.

\section{Atom number distribution in the optical lattice}

\begin{figure}[t]
\begin{center}
\includegraphics{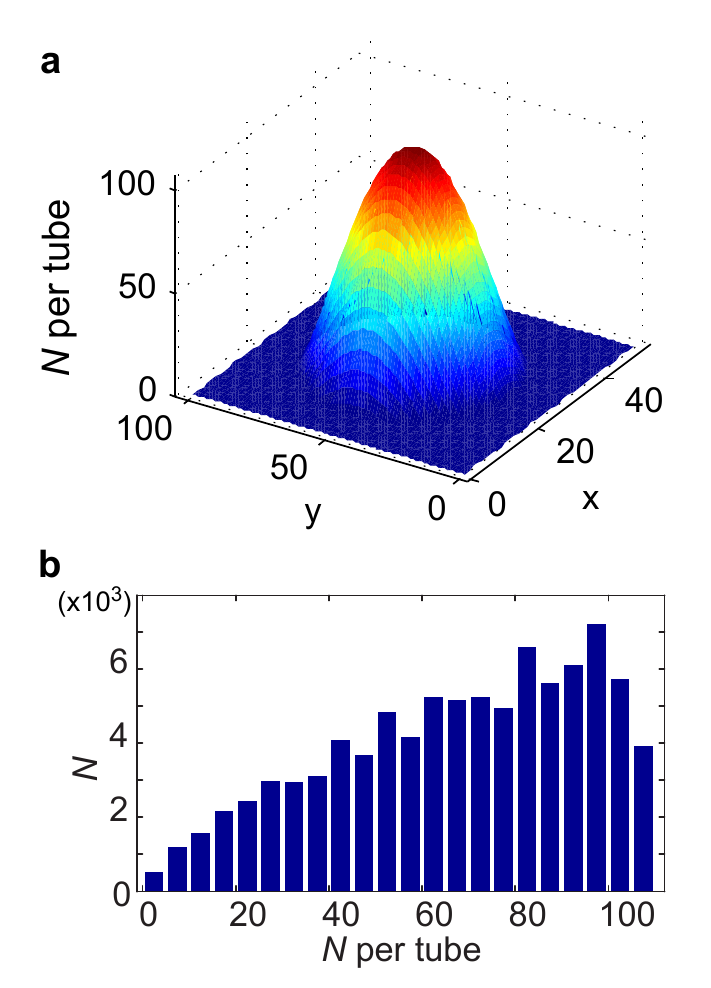}
\end{center}
\caption{
Atom number distribution in the system. \textbf{a} Atom number distribution in the 2D lattice right after the initial loading. For our parameters and the total number of particles ($9\times10^4$), we get about 108 atoms in the central tube of the system and the occupation decreases with distance from the center. The mean value is 43 atoms per tube. \textbf{b} Histogram showing the total number of atoms residing in tubes with given occupation numbers.
}
\label{fig:atomNo}
\end{figure}

In our experiment, the presence of the magnetic trap and the Gaussian profile of the lattice beams leads to an overall harmonic confinement and correspondingly inhomogeneous distribution of atom numbers in the tubes. The trap frequencies right after ramping up the initial 2D optical lattice are $\omega_x = 2\pi \times45\,{\rm Hz}$, $\omega_y = 2\pi\times40\,{\rm Hz}$ and $\omega_z = 2\pi\times59\,{\rm Hz}$, respectively. Using the effective description of 1D-gases (see e.g. \cite{Petrov00-s}) and an $s$-wave scattering length of $a_s = 5.34\,{\rm nm}$, we simulate the atom number distribution employing a local density approximation. This results from the loading sequence described in the Methods section of the main article. For our typical number of $9\times10^4$ atoms, we find a maximum of 108 particles per tube in the center (see Fig.~\ref{fig:atomNo}) right after the initial rampup of the 2D lattice.

\section{The bosonic ladder}

Here we provide details for the theoretical description of the coupled two-tube system with a focus on the Landau-Zener (LZ) passage discussed in the main article. The details given here cover two different aspects of the coupled Bose-liquids: (i) The energetics which can be captured in a mean-field description. The mean-field solution allows us to map the many-body system to an effective two-state problem discussed by Wu {\em et al.} \cite{Wu00-s,Liu02-s}. We show how to extract an adiabaticity criterion from this mapping. (ii) The dynamics of the collective excitations of the Bose liquids, which provide an effective bath for the LZ problem. In particular, we describe how to capture the decay processes responsible for breakdown of adiabaticity in the inverse sweep.

\subsection{Non-linear Landau-Zener problem}
\label{sec:mapping}

Before we analyze the Hubbard Hamiltonian on the two-leg ladder we review the physics of a simpler non-linear two-state model \cite{Wu00-s,Liu02-s} described by the classical Hamiltonian:
\begin{align}
\label{eqn:wu-ham}
{\mathcal H}[\psi_{L},\psi_{R}]&=-2J\psi_{L}\psi_{R} + \Delta(|\psi_{L}|^{2}-|\psi_{R}|^{2})
\\
\nonumber
&\quad+U (|\psi_{L}|^{4}+|\psi_{R}|^{4}).
\end{align}
Here, $\Delta$ is the detuning which will be ramped as $\Delta=\alpha\cdot t$. Note, that we consider $U>0$. The two amplitudes $\psi_{R}$ and $\psi_{L}$ obey $|\psi_{R}|^{2}+|\psi_{L}|^{2}=1$.

\begin{figure}[t]
\begin{center}
\includegraphics{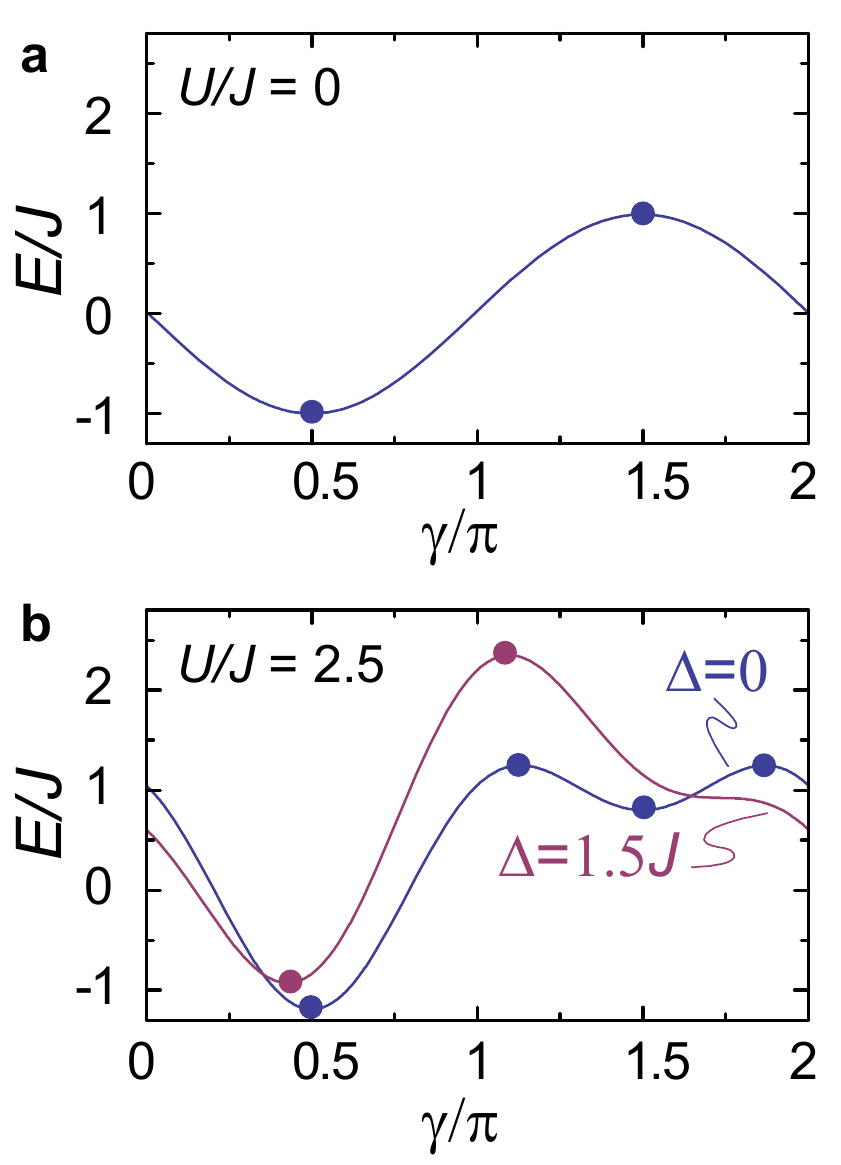}
\end{center}
\caption{
The energy of the non-linear Hamiltonian (\ref{eqn:wu-ham}) as a function of the angle $\gamma$. The extrema mark the (meta-)stable solutions. \textbf{a} For no non-linearity ($\eta=0$), only two states are stable (dots). \textbf{b} Depending on the detuning either four ($\Delta=0$) or two ($\Delta=1.5\,J$) states appear close to resonance. %
}
\label{fig:angleproblem}
\end{figure}

For $\psi_{R,L}$ real we parametrize $\psi_{R}=\sin(\gamma/2)$ and $\psi_{L}=\cos(\gamma/2)$. Fig.~\ref{fig:angleproblem} shows the energy landscape as a function of the  angle $\gamma$.  For sufficiently strong non-linearity one finds four extrema of the energy near zero bias and only two further away. This is essentially the loop-structure discussed in  \cite{Wu00-s}. The loop structure appears beyond a critical value of the non-linearity , i.e. $\eta = U/J>1$. For $\eta<1$ and $\Delta=0$, there are the two extrema corresponding to the bonding and anti-bonding state of the non-interacting problem at $\gamma=\pi/2$ and $\gamma=3\pi/2$, respectively. For $\eta>1$ the non-linearity is strong enough to give rise to four different extrema at zero detuning. When going away from $\Delta=0$, the detuning introduces a tilt and two of the extrema eventually merge and disappear. The resulting energy level structure is given in Fig. 3{\bf b} of the main article.

The condensates $\psi_{L,R}$ obey the (Gross-Pitaevskii) dynamics
\begin{equation}
\label{eqn:schdyn}
i\hbar \partial_{t} \psi_{L} = \partial_{\psi_{L}^{*}}{\mathcal H}(\psi_{L},\psi_{R}),
\end{equation}
prescribed by the Hamiltonian (\ref{eqn:wu-ham}) together with the Poisson brackets $\{\psi_\alpha,\psi^*_\alpha\}=1$. Based on this dynamics it is possible to compute a transition  probability, using the adiabatic perturbation theory of classical mechanics \cite{Landau1-adiabatic-s}. This yields \cite{Liu02-s} $P\propto 1-\exp(-\alpha_{c}/\alpha)$ with $\alpha_{c}=2\pi J^{2} q$  and
\begin{equation}
\label{GP2}
q=\frac{4}{\pi}\int_{0}^{\eta^{-1/3}}\!\!\!\!dx\,
(1+x^{2})^{1/4}\bigl[(1+x^{2})^{-3/2}+\eta\bigr]^{3/2}.
\end{equation}
Note that $q=1$ for $\eta=0$, in accordance with the exact result.

\subsection{From a Bose-Hubbard ladder to two-mode dynamics}

In the main text we described a mapping from the two chain Hubbard model to the energetics of the non-linear two-mode LZ problem by employing a mean-field approximation for the inter-tube coherence. Here we will show that the dynamics is of the form (\ref{eqn:schdyn}) and derive the effective non-linearity $\eta$ in this case.

As described in the text, the key is to treat the inter-tube coherence in a single-mode  approximation taking a uniform inter-tube wave-function for the condensate $\hat a\yd=\psi_L \hat b\yd_L+\psi_R \hat b\yd_R$. The resulting Hamiltonian $H_{\ssm eff}$ given by Eq. (6) in the Methods section is not exactly solvable.  To make further progress we apply a mean-field (MF) approximation for both the axial and inter-tube modes. This of course misses the subtle (power-law) long range correlations in the one-dimensional Bose liquid, however it is not a bad approximation for the energetics.

The MF theory can be easily formulated as a variational approximation with $\ket{\Psi}$ taken to be a site factorizable wave-function, which depends on the inter-tube coherences $\psi_L,\psi_R$ among other parameters. The time evolution, or the equations of motion for the variational parameters, are obtained from the time dependent variational principle \cite{Jackiw79-s,Huber08-s}
\begin{equation}
\mathcal{L}_{\ssm eff}= \langle \Psi | i\hbar\partial_{t} - H | \Psi\rangle,
\end{equation}
where $H$ is the original ladder Hamiltonian (5).  In the weak coupling limit $U\ll J_\parallel n$, the variational wave-function can be taken to be the coherent state
\begin{equation}
| \Psi_{\psi_{L,R}}\rangle =
e^{-|\Phi|^{2}+\Phi(\psi_{L}\hat b_{L}^{\dag}+\psi_{R}\hat b_{R}^{\dag})}|{\rm vac}\rangle,
\end{equation}
which gives rise to the two-mode Gross-Pitaevskii dynamics
\begin{align}
\nn
\mathcal{L}_{\ssm eff}&=
i\hbar(
\psi_{L}\dot\psi_{L}^{*}+
\psi_{R}\dot\psi_{R}^{*})
\\
\nn
&\quad
-J(\psi^*_L\psi_R+{\rm c.c.})
+\Delta
(|\psi_L|^2-|\psi_R|^2)\\
\label{eqn:gpe-eff}
&\quad+U\Phi^{2}(|\psi_{L}|^{4}+|\psi_{R}|^{4})-2J_{\parallel}\Phi^{2}
\end{align}
with non-linearity $\eta=U|\Phi|^2/J$. The absence of dynamics in the axial degree of freedom is a result of constraining the variational states to a uniform condensate $\Phi$. In reality there are low-energy axial phonon excitations, the consequences of which will be discussed in the last section.

For strong interactions we resort to the Gutzwiller MF theory, which consists in using the site-factorizable variational states $ \ket{\Psi_{\psi_{L,R}}}=\prod_i f_m\ket{m}_i$. Here $\ket{m}_i$ denotes a state with $m$ $\hat a_{i}$-bosons on site $i$. Very close to the Mott transition and in the Mott phase we may truncate to only three occupation states, so that
\begin{align}
\nn
|\Psi_{\psi_{L,R}}\rangle&=\prod_i
\left[\cos(\theta/2) |1\rangle_i + \frac{\sin(\theta/2)e^{i\phi}}{\sqrt{2}}
\left(|0\rangle_i +  |2\rangle_i\right)\right]\\
\nn
&=\prod_i \bigg[\cos(\theta/2)
\big[\psi_{L}\hat b_{L}^{\dag}+\psi_{R} \hat b_{R}^{\dag}\big]\\
\nn
&\quad+
\frac{\sin(\theta/2)e^{i\phi}}{\sqrt{2}}
\bigg\{1+
\frac{\big[\psi_{L}\hat b_{L}^{\dag}+\psi_{R} \hat b_{R}^{\dag}\big]^{2}}{\sqrt{2}}\bigg\}\bigg]|{\rm vac}\rangle.
\end{align}
We evaluate the effective Lagrangian $\mathcal{L}_{\ssm eff}$ to leading order in small $\Phi= \langle \Psi| \hat a_{i} | \Psi\rangle$ to obtain
\begin{align}
\label{eqn:line1}
\mathcal{L}_{\ssm eff}
&=i\hbar (\psi_{L}^{\phantom{*}}\dot\psi_{L}^{*}+\psi_{R}^{\phantom{*}}\dot\psi_{R}^{*}) \\
&\quad-J(\psi^*_L\psi_R^{\phantom{*}}+{\rm c.c.})+\Delta
(|\psi_L|^2-|\psi_R|^2)\\
\label{eqn:line3}
&\quad+U\mathcal{C}\Phi^{2}(|\psi_{L}|^{4}+|\psi_{R}|^{4})
-(3+2\sqrt{2})J_{\parallel}\Phi^{2}
\\
&\quad+
\frac{\hbar}{2}\mathcal{C} \Phi^{2}\dot\phi-\frac{i\hbar}{2}\mathcal{C}|\psi_{L}|^{2}|\psi_{R}|^{2}\dot\Phi \Phi+4J_{\parallel}\Phi^{2}\phi^{2}
\\
&\quad+
\Phi^{4}
\left(
\frac{J_{\parallel}(3+2 \sqrt{2})}{6}-\frac{U\mathcal{C}(|\psi_{L}|^{4}+|\psi_{R}|^{4})}{12}\right).
\label{eqn:lastline}
\end{align}
The numerical factor $\mathcal{C}\lesssim 1$ results from the truncation.

In the lines (\ref{eqn:line1})--(\ref{eqn:line3})  we recognize again the non linear two mode-dynamics for $\psi_{L,R}$ with non-linearity $\eta=U\Phi^2/J$.  In general this dynamics is coupled to the uniform dynamics of $\Phi$ which correspond to a gapped condensate amplitude oscillation about the average value ${\bar\Phi}$ dictated by the $\Phi^4$ theory [lines (\ref{eqn:line3})--(\ref{eqn:lastline})] and depends on the distance from the Mott phase. For slow LZ dynamics we can therefore set $\Phi$ and $\f$ to their instantaneous variationally determined values $\bar\Phi$ and $\f=0$.

To obtain the estimate for the critical ramp rate $\a_c$ we compute the condensate fraction $\Phi$ at $\D=0$ for given $U$, $J_{\parallel}$, and $J$. This gives the characteristic non-linearity parameter  $\eta$ in the effective model. We then use $\eta$ to calculate the critical $\alpha_{c}$ via (\ref{GP2}). The result is shown in the insets of Fig. 2{\bf c} and Fig.~2{\bf d} in the main article.

\subsection{Phase diagram and maximal loop}

In this section we derive the phase diagram of the meta-stable Bose liquid in the upper state and determine the point of maximal non-linearity for each given inter-tube coupling $J$ as a function of the axial-lattice depth. This could be done, in principle, using the inter-tube MF approximation described above.  However, to get the complete phase diagram including the region of very small $J$ it is better to use a MF approximation which treats the inter-tube coupling exactly, keeping both the bonding and anti-bonding inter-tube states. This is done using the following family of site factorizable variational states $|\Psi\rangle=\prod_{i}|\psi_{i}\rangle$ with
\begin{align}
\nonumber
|\psi_{i}\rangle &=
\cos \theta
\Bigg\{
 \cos\frac{\chi}{2} |10\rangle_{i} \!+\! \sin\frac{\chi}{2}|01\rangle_{i}
\Bigg\}
+\frac{e^{i\phi}\sin \theta}{\sqrt{2}}\times
\\
\label{eqn:wavefunction}
&\!\!\!\!\!\!\!\!\!\!\!\!\Bigg\{
|00\rangle_{i} \!+\!
\bigg[
\sin \iota
\bigg(\cos\frac{\chi}{2} |20\rangle_{i} \!+\! \sin\frac{\chi}{2} |02\rangle_{i}\bigg)
\!+\!\cos\iota| 11\rangle_{i}
\bigg]
\Bigg\}.
\end{align}
The Fock states $|\alpha\beta\rangle_{i}$ denote occupation states at site $i$ in the left ($\alpha$) and right ($\beta$) tube. This wave function, improves on the inter-tube MF, described in the previous section, by allowing for combinations of the states $|\alpha\beta\rangle_{i}$, which cannot be written using the operators $a_{i}$ alone.
\begin{figure}[t]
\begin{center}
\includegraphics{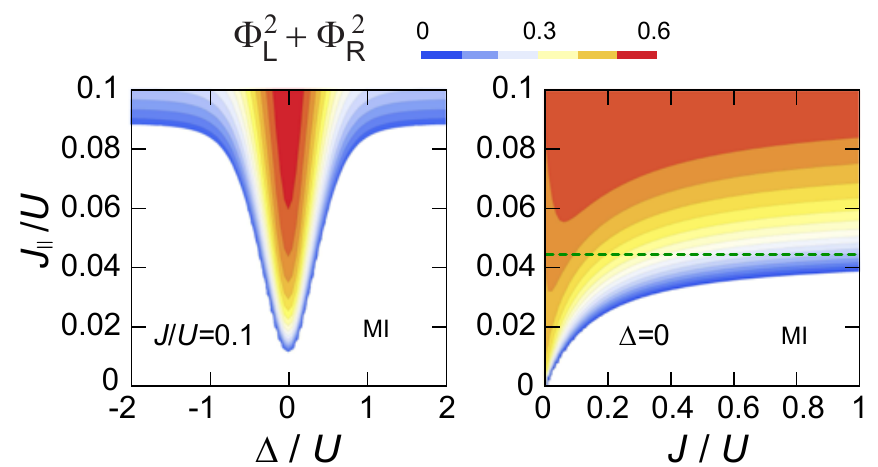}
\end{center}
\caption{
Adiabatic phase diagram within the variational wave function approach (\ref{eqn:wavefunction}). White denotes the Mott insulator, the color represents the expectation value of the kinetic energy. In the right panel we show in comparison the critical value of $J_{\parallel}$ obtained from the simpler mean-field approximation described by Eq. (\ref{eqn:wu-ham}) as a dashed line.
}
\label{fig:pd}
\end{figure}

The phase diagram computed using the states (\ref{eqn:wavefunction}) is described as follows. First it is clear that for large detuning ($\Delta\gg J,U$) one can neglect the lower tube (for the inverse sweep) and the phase diagram is that of a single chain.  To understand the more interesting regime near resonance, we focus on the case $\Delta=0$ for simplicity.  The inter-tube hopping acts to reduce the effect of interactions, significantly shrinking the Mott phase. Fig.~\ref{fig:pd}\textbf{b} displays a map of the superfluid order parameter $|\Phi_L|^2+|\Phi_R|^2$ in the space of $J_\parallel/U$ versus $J/U$. Fig.~\ref{fig:pd}\textbf{a} shows the phase diagram for the case $\D=0$.  These results are in good qualitative agreement with Ref.~\cite{danshita07-s}. To get more precise results, Danshita {\em et al.} resorted to numerical DMRG calculations.

The results shown in Fig.~\ref{fig:pd}\textbf{b} are used in Fig.~5{\bf b}. (i) The phase boundary shown in Fig. 5{\bf b} is defined by the line where $\Phi\rightarrow 0$. (ii) We took a contour of equal order parameter $\Phi=\Phi_{0}$ close to its maximal value as an estimate for the location of the maximal loop structure.

\bigskip

\subsection{Relaxation due to axial phonons of the Bose liquid}

The crucial ingredient leading to the failure of adiabaticity in the inverse sweep, which was missing in the above MF considerations, is the existence of phonon modes in the 1D quantum liquid. These modes facilitate the decay of the inter-tube condensate wave-function from the meta-stable upper to the lower state via emission of phonons.  In this section we provide the details of the effective Hamiltonian used to theoretically account for this effect.

We start by deriving the coupling of the two adiabatic modes. For the discussion of the quasi-condensate we only used the state $\hat a_{i}^{\dag}=\psi_{L} \hat b_{iL}^{\dag}+\psi_{R} \hat b_{iR}^{\dag}$. To describe the coupling of the two adiabatic modes we also need  the orthogonal state $\hat b_{i}^{\dag}=-\psi_{R} \hat b_{iL}^{\dag}+\psi_{L} \hat b_{iR}^{\dag}$.  As long as the loop structure does not develop, i.e. for weak non-linearity $\eta\ll 1$, the two amplitudes $\psi_{L,R}$ are approximately given by the solution of the single-particle double-well Hamiltonian
\begin{equation*}
\psi_{L,R}=\sqrt{\half\left(1\pm \frac{\D}{\D_{\ssm eff}}\right)},
\end{equation*}
where $\Delta_{\ssm eff}\equiv \sqrt{4J^{2}+\Delta^{2}}$. This is valid for example in and close to the Mott phase.

We can now express the interactions in terms of the operators $\hat a_{i}$ and $\hat b_{i}$ to obtain the coupling terms between the modes:
\begin{widetext}
\begin{equation}
\label{eqn:rotint}
H_{\ssm int} = \frac{U}{2}\sum_{i}
\biggl\{
\lambda_{0} (\hat \rho_{ai}^{2}+\hat \rho_{bi}^{2}) +
\lambda_{12} \hat\rho_{ai}\hat\rho_{bi} +
\lambda_{1}
(\hat\rho_{ai}-\hat\rho_{bi})
\bigl[
\hat a_{i}^{\dag}\hat b_{i}^{\pdag}+{\rm H.c}
\bigr]
+
\lambda_{2}
\bigl[
\hat a_{i}^{\dag} \hat a_{i}^{\dag}\hat b_{i}^{\pdag}\hat b_{i}^{\pdag}+{\rm H.c}
\bigr]
\biggr\}.
\end{equation}
We defined
\begin{equation*}
\hat \rho_{ai}=\hat a_{i}^{\dag}\hat a_{i}^{\pdag},\quad
\hat \rho_{bi}=\hat b_{i}^{\dag}\hat b_{i}^{\pdag},\quad
\lambda_{0}=1-\frac{2J^{2}}{\Delta_{\ssm eff}^{2}},\quad
\lambda_{12}=\frac{8J^{2}}{\Delta_{\ssm eff}^{2}},\quad
\lambda_{1}=\frac{2\Delta J}{\Delta_{\ssm eff}^{2}},\quad
\lambda_{2}=\frac{4J^{2}}{\Delta_{\ssm eff}^{2}}.
\end{equation*}
\end{widetext}
The coupling Hamiltonian (\ref{eqn:rotint}) contains an intra-mode interaction $\propto \lambda_{0}$, an inter-mode interaction $\propto \lambda_{12}$, and particle and pair transfer terms $\lambda_{1}$, and $\lambda_{2}$, respectively.

To understand the effect of the coupling on the Landau-Zener dynamics in the limit of slow sweeps, we make use of the effective long wavelength (bosonized) description of the quantum liquid in the adiabatic upper mode. This is given by the Sine-Gordon model \cite{giamarchi04-s}:
\bea
H_{a} &=& \frac{\hbar v_{s}}{2\pi} \int dx\, K |\partial_{x}\hat\varphi(x)|^{2}+\frac{1}{K}
|\partial_{x}\hat\vartheta(x)|^{2}\nn\\
&&-g\int dx\, \cos\left[2\hat\vartheta(x)\right].
\label{Ha}
\eea
Here $v_s\approx l\sqrt{U J_{\parallel}n}$ is the sound velocity with $l$ the lattice constant.  The coupling constant $g$ and the Luttinger parameter $K$ are in general  phenomenological parameters. In the superfluid phase $K>2$ and it decreases with increasing lattice depth. The transition to a Mott insulator occurs at the universal value of $K=2$, below which the cosine term becomes relevant, giving rise to the Mott gap $\D_{\ssm MI}$.  Note that the long wavelength theory is valid at energies up to the UV-cutoff $\w_0\sim \sqrt{U J_{\parallel}n}$ above the upper inter-tube state. At the Mott transition $\w_0\sim U$.

The dynamics of a particle, which decays to the lower inter-tube mode is described as a free particle at the bottom of an empty band
\be
\label{Hb}
H_{b} = \int \frac{dk}{2\pi} \frac{\hbar^{2}k^{2}}{2m^{*}} {\hat \psi}_{b,k}^{\dag}{\hat\psi}_{b,k}^{\pdag},
\ee
where $m^{*}\approx \hbar^{2}/J_{\parallel}l^{2}$ is the effective mass.

The coupling Hamiltonian (\ref{eqn:rotint}) can also be written in terms of the long-wavelength (bosonized) degrees of freedom. The most relevant term due to single particle tunneling maps to:
\be
H_1= A_1 U \lambda_1 \rho_a^{3/2}\int  dx\, e^{i \hat\f(x)}\hat \psi_b(x)+{\rm H.c.}
\label{H1}
\ee
Similarly for the two particle tunneling it is
\be
H_2= A_2  U \lambda_2 \rho_a\int dx\, e^{i2 \hat\f(x)}\hat\psi_b(x)\hat\psi_b(x)+{\rm H.c}.
\label{H2}
\ee
Here, $A_1$ and $A_2$ are dimensionless non-universal constants of order $1$.  Finally, we must also consider the cross-interaction term $\lambda_{12}$, between the two modes.  This gives rise to a ``final-state interaction'' between the quantum liquid in state $a$ and the first particles transferred to the mode $b$. In this respect, the particles in the initially empty mode represent mobile scattering centers (impurities) for the full mode.  It is known that {\em mobile} impurities do not lead to back-scattering \cite{Castro-Neto96-s} at low energies. For this reason we can write the cross-interaction as
\begin{equation}
\label{eqn:forward}
H_{12} = A_{12}U\lambda_{12}{\pi}
\int dx\, \hat\rho_{b}(x)\partial_{x}\hat\vartheta(x).
\end{equation}
Again $A_{12}$ is a constant of order $1$. The coupling (\ref{eqn:forward}) can be removed by a unitary transformation \cite{fiete09-s}
\begin{equation}
\hat U=\exp\left[-i\delta_{s}\int dx' \hat\rho_{b}(x')\hat\varphi(x')\right],
\end{equation}
with $\delta_{s}=A_{12}U\lambda_{12}K/\hbar v_{s}\pi$. This transformation leaves the Hamiltonians (\ref{Ha}) and (\ref{Hb}) invariant but transforms the operators as
\begin{equation}
\hat\psi_{a}^{(\dag)}(x)\rightarrow \hat\psi_{a}^{(\dag)}(x)\, \quad
\hat\psi_{b}^{(\dag)}(x)\rightarrow \hat\psi_{b}^{(\dag)}(x)e^{\mp i\delta_{s}\hat\phi(x)}.
\end{equation}

We can now derive the decay rates $\Gamma_{1(2)}(\Delta,K)$ from the state $a$ to $b$ due to the tunneling terms (\ref{H1}) and (\ref{H2}) within a Fermi-golden-rule approach. For this we have to calculate the correlation functions
\begin{align*}
\mathcal{C}_{1}&=
\langle \hat\psi_{b}^{\dag}(x)\hat\psi_{a}^{\dag}(x)\hat\psi_{b}(0)\hat\psi_{a}(0)\rangle, \\
\mathcal{C}_{2}&=
\langle \hat\psi_{b}^{\dag}(x) \hat\psi_{b}^{\dag}(x)\hat\psi_{a}^{\dag}(x)\hat\psi_{a}^{\dag}(x)\hat\psi_{b}
(0)\hat\psi_{b}(0)\hat\psi_{a}(0)\hat\psi_{a}(0)\rangle,
\end{align*}
for the $\lambda_{1}$ and $\lambda_{2}$ processes, respectively. Due to the ``final-state-interaction'', the operators $\hat\psi_{a}(x)$ acquire an extra exponent $i\delta_{s}\hat\varphi(x)$ which effectively changes the scaling dimension of the correlation functions $C_{1}$ and $C_{2}$. We can absorb this by rescaling the Luttinger parameter
\begin{equation}
K^{*}=\frac{K}{(1+\delta_{s})^{2}}.
\end{equation}
Note that this rescaling is only relevant for the correlation functions $\mathcal{C}_{1}$ and $\mathcal{C}_{2}$. For the description of processes in the isolated mode $a$, e.g., the discussion of instabilities of the Luttinger liquid, the relevant parameter is still the original $K$. The rates $\Gamma_{1(2)}(\Delta,K)$ given in the main article are now obtained  by standard manipulations \cite{Huber09b-s}.

\end{document}